\documentclass[aps,prl,twocolumn,groupedaddress,showpacs,letterpaper,floatfix,10pt]{revtex4-1}
\usepackage{amsmath,amssymb,graphicx,hyperref}
\hypersetup{pdfnewwindow=true, colorlinks=true, linkcolor=blue, anchorcolor=blue,
  citecolor=blue, filecolor=blue, menucolor=blue, urlcolor=blue}

\def\k{{\bf k}}

\def\avg#1{\langle #1 \rangle}

\def\phii{\vec{\phi}_\text{i}}
\def\phim{\vec{\phi}_\text{m}}
\def\di{d_\text{i}}
\def\dm{d_\text{m}}
\def\ni{n_\text{i}}
\def\nm{n_\text{m}}
\def\iwn{i\omega_n}

\def\iw{{i\omega}}
\def\cM{{\cal M}}
\def\cI{{\cal I}}
\def\cT{{\cal T}}

\begin{document}

\title{Metal-Mott insulator interfaces}

\author{Juho Lee}
\author{Chuck-Hou Yee}
\affiliation{Dept. of Physics \& Astronomy, Rutgers, The State University of New Jersey, Piscataway, NJ 08854, USA}

\date{\today}

\begin{abstract}
  Motivated by the direct observation of electronic phase separation in
  first-order Mott transitions, we model the interface between the
  thermodynamically coexisting metal and Mott insulator. We show how to model
  the required slab geometry and extract the electronic spectra. We construct
  an effective Landau free energy and compute the variation of its parameters
  across the phase diagram. Finally, using a linear mixture of the density and
  double-occupancy, we identify a natural Ising order parameter which unifies
  the treatment of the bandwidth and filling controlled Mott transitions.
\end{abstract}


\maketitle

\section{Introduction}

First-order transitions exhibit phase separation, and the real-space structure
of the interface between the two thermodynamic phases contains information
about the free energy functional~\cite{Binder1987}. Specifically, the thickness
of the interface allows direct access to the ratio of the potential to kinetic
energy terms in the free energy, which is related to the barrier height between
the two minima of the double-well. A widely observed first-order transition in
solid state systems is the Mott transition (reviewed in~\cite{Imada1998} for a
large class of materials). Here, temperature, pressure or chemical doping
drives a transition between a metal and a Mott insulator, a state where
electrons cannot conduct due to the large ratio of the local Coulomb repulsion
relative to the kinetic energy. While phase separation at the Mott transition
is theoretically well-studied~\cite{Visscher1974, Emery1990, Gehlhoff1996,
  White2000, Galanakis2011, Yee2015a}, the interface between the
thermodynamically coexisting metal and Mott insulator is not. The recent
development of experimental probes with nanometer-scale spatial
resolution~\cite{Hanaguri2004, Qazilbash2007, Kohsaka2012} has allowed the
direct observation of the real-space structure of these interfaces.

As a first step towards characterizing the metal-Mott interface, we compute the
real-space structure of the interfaces for the canonical example of a
correlated system, the single-band Hubbard model. We use techniques in the
spirit of work on correlated surfaces~\cite{Schwieger2003, Ishida2006} and
heterostructures~\cite{Freericks2004, Helmes2008, Zenia2009, Borghi2010,
  Bakalov2016}. We extract the evolution of the density, double-occupancy and
spectral features across the interface, allowing us to determine the parameters
of the underlying free energy across the phase diagram.

\begin{figure}
  \includegraphics[width=\columnwidth]{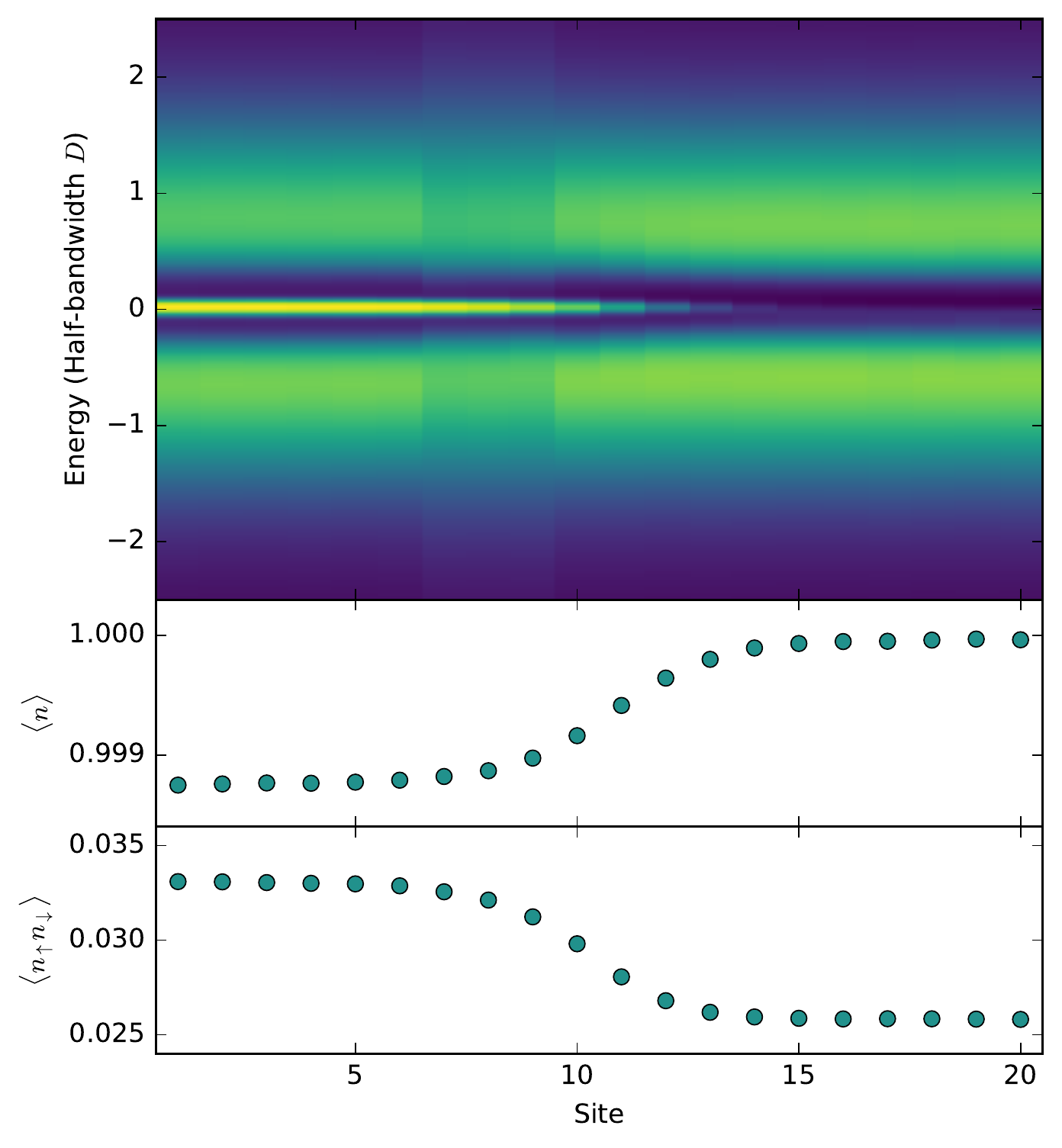}
  \caption{Evolution of the local spectra (top panel), density (middle) and
    double occupancy (bottom) across the interface between a correlated metal
    (left edge) and a Mott insulator (right edge). Clearly visible is the
    transfer of spectral weight from the low-energy quasiparticles to the
    Hubbard bands as we spatially traverse the interface. We have chosen
    parameters of the Hubbard model where the transition from the insulator is
    to a hole-doped metal: $\mu = 0.95(U/2)$, $U = 1.97D$ and $T = 0.01D$,
    where $D = 6t$ is the half-bandwidth.}
  \label{fig:spectra}
\end{figure}

\section{Landau free energy}

The Mott transition can be tuned by two parameters besides temperature: the
chemical potential $\mu$ and correlation strength $U$. At half-filling,
extensive work has shown the first-order transition is analogous to the
liquid-gas transition, placing the Mott transition within the Ising
universality class~\cite{Castellani1979, Kotliar2000, Onoda2003, Limelette2003,
  Kagawa2005, Papanikolaou2008, Semon2012, Vucicevic2013}. In this work, we
extend the construction away from half-filling into the $\mu$-$U$
plane~\cite{Werner2007}. Since we are interested in the metal-Mott interface,
we work at temperatures below the critical point to construct our Landau
theory.

We choose our fields to be the quantities conjugate to the external parameters
$(\mu, U)$, namely the density $n = \avg{n}$ and double occupancy $d =
\avg{n_\uparrow n_\downarrow}$, a construction hinted at in~\cite{Imada2005}.
The transition between the metal and paramagnetic Mott insulator does not break
any symmetries~\cite{McWhan1970a, Imada1998}, so the terms in the free energy
functional $\mathcal{F}[n,d]$ are unconstrained. The free energy generically
will have one global minimum, and should a transition exist, it will occur via
the switching between two discrete minima as no symmetry forces a locus of
states to simultaneously lower in energy. We will explicitly construct the
scalar order parameter in the following


\begin{figure}
  \includegraphics[width=\columnwidth]{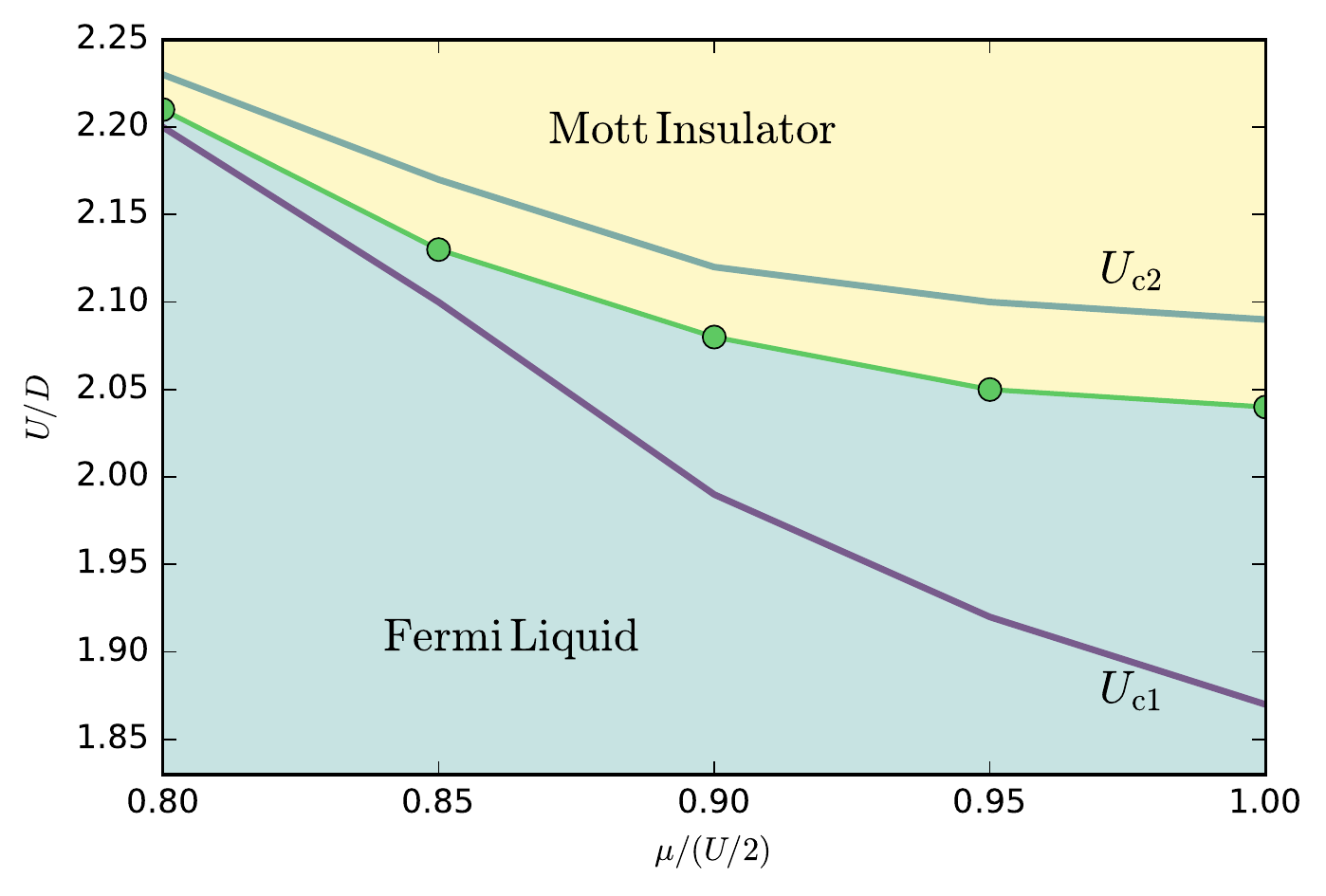}
  \caption{Generic Mott phase diagram, as realized by the single-band Hubbard
    model on a cubic lattice at $T = 0.005D$, where $D = 6t$ is the
    half-bandwidth. Dots label values of $(\mu,U)$ lying on first-order Mott
    transition line used for interface calculations. Lines mark the spinodals
    $U_{\text{c}1}$ and $U_{\text{c}2}$ where the insulating and metallic
    solutions respectively vanish. The diagram is symmetric about $\mu/(U/2) =
    1$.}
  \label{fig:phase-diag}
\end{figure}

Along the Mott transition line in the $\mu$-$U$ plane, the two minima will have
the same energy. To facilitate analytic calculation, we take the two minima to
be symmetric, an assumption certainly not justified by symmetry, but which will
prove to be a good approximation. Writing the fields as $\vec{\phi} = (n,d)$,
the free energy functional takes a double-well form,
\begin{equation}
  \mathcal{F}[\vec{\phi}] = \frac{1}{2} D(\nabla\vec{\phi})^2 + \\ \lambda
  (\vec{\phi} - \vec{\phi}_\text{i})^2 (\vec{\phi} - \vec{\phi}_\text{m})^2,
  \label{eqn:free}
\end{equation}
where $\phii = (\ni,\di)$ and $\phim = (\nm,\dm)$ are the insulating and
metallic minima. A note on units: we work on a discrete lattice to easily
connect with computation and set the lattice spacing $a = 1$. Thus the
gradient is understood to be discrete $\nabla\vec{\phi}_j \sim \vec{\phi}_{j+1}
- \vec{\phi}_j$, where $j$ is the lattice site, the free energy $F = \sum_j
\mathcal{F}[\vec{\phi}_j]$, and both $\lambda$ and $D$ have units of energy. We
choose $D$ to be the half-bandwidth and omit an overall (dimensionless)
normalization to the free energy.

A domain wall is given by the standard solution used, e.g. in the theory of
instantons~\cite{Vainshtein1982},
\begin{equation}
  \vec{\phi}(x_j) = \frac{\phim+\phii}{2} + \frac{\phim-\phii}{2}\tanh\left(\frac{x_j-x_0}{2l}\right)
  \label{eqn:soliton}
\end{equation}
where $x_j$ is the coordinate of the $j$th site and the wall thickness is
$l^{-2} = 2(\lambda/D)(\phim-\phii)^2$. Note the fields $\vec{\phi}$ do not
transform as a vector and the notation is for convenience. Determining the
dependence of $\phii$, $\phim$ and $\lambda/D$ on $(U,T)$ requires microscopic
modeling.


\section{Modeling the Interface}

The Hubbard hamiltonian is the ``standard model'' of correlated electrons. Its
two terms describe the competition between kinetic energy, which delocalizes
electrons to promote metallic behavior, and mutual electron repulsion, which
tends to localize electrons onto sites and drive the transition to a Mott
insulator. We work with the simplest one-band case on a cubic lattice,
\begin{equation}
  H = \sum_{\k\sigma} (\epsilon_k - \mu) n_{\k\sigma} \\
  + U \sum_j n_{j\uparrow} n_{j\downarrow},
\end{equation}
where we take $\epsilon_\k = -2t(\cos k_x + \cos k_y + \cos k_z)$ and use the
half bandwidth $D = 6t$ as the unit of energy in all the following. We will
index the sites by $j = (n_1, n_2, n_3)$ in the following. Ignoring ordered
phases, which is a reasonable assumption at intermediate temperatures or in the
presence of frustration, the phase diagram generically consists of a Mott
insulating region for large $U$ and a range of $\mu$ corresponding to
half-filling, and a Fermi liquid everywhere else. To find the first-order
transition line, we use standard single-site dynamical mean-field theory (DMFT)
~\cite{Metzner1989, Georges1992, Georges1996} with a continuous-time quantum
monte carlo (CTQMC) hybridization expansion impurity solver~\cite{Werner2006,
  Haule2007b, Gull2011}. The phase diagram at $T = 0.005D$ is plotted in
Fig.~\ref{fig:phase-diag}, along with the two spinodals $U_{\text{c}1}$ and
$U_{\text{c}2}$ between which both solutions exist.

\begin{figure}
  \includegraphics[width=\columnwidth]{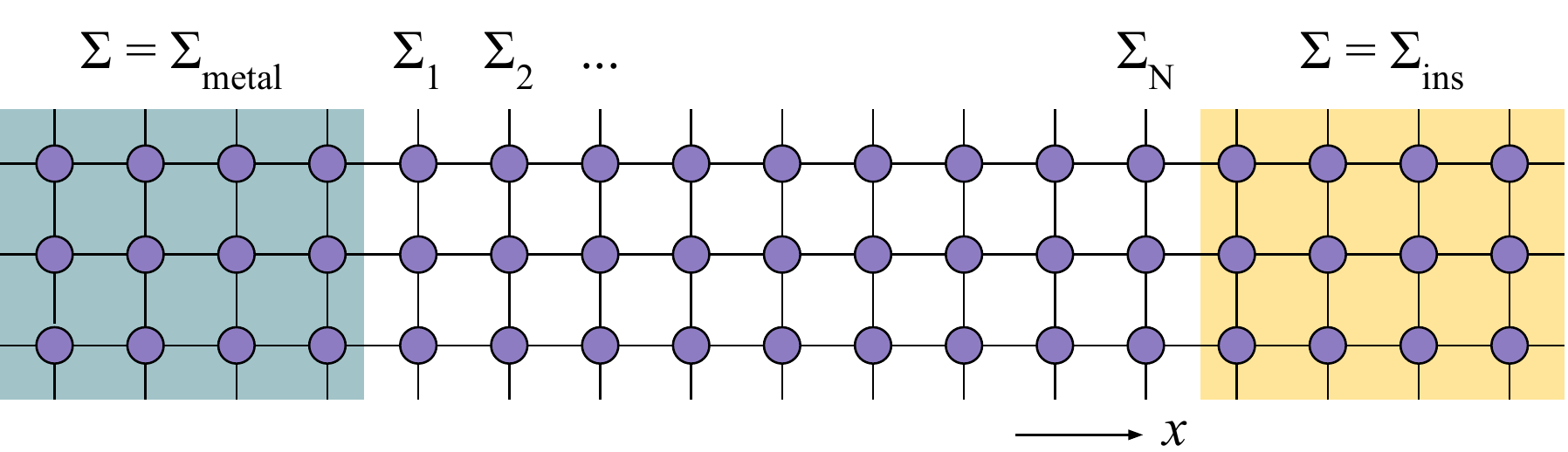}
  \caption{Geometry used to model the metal-Mott interface. The transition
    region, described by site-dependent self-energies $\Sigma_i$, is sandwiched
    between a semi-infinite bulk Fermi liquid and Mott insulator by fixing the
    bulk self-energies to $\Sigma_\text{metal}$ and $\Sigma_\text{ins}$ on the
    left and right. We assume translational invariance in the $y$ and $z$
    directions.}
  \label{fig:slab}
\end{figure}

To model the interface in the coexistence regime, we fix our parameters to a
point on the first-order line (dots in Fig.~\ref{fig:phase-diag}), then
partition the lattice into three regions along the $x$-axis (see
Fig.~\ref{fig:slab}): metal ($n_1 \leq 0$), insulator ($n_1 \geq N+1$), and a
\emph{transition} region ($1 \leq n_1 \leq N$). Here $n_1$ is the site index
along the $x$-axis and we take $N = 20$ large enough to capture the interface.
We perform an inhomogenous DMFT calculation by setting the self-energy of the
lattice $\Sigma_{n_1 n_1'} = \delta_{n_1 n_1'} \Sigma_{n_1}$ to
\begin{align}
  \Sigma_{n_1} = \begin{cases}
    \Sigma_\text{metal} & n_1 \leq 0 \\
    \Sigma_{n_1}        & 1 \leq n_1 \leq N \\
    \Sigma_\text{ins}   & n_1 \geq N+1 \\
  \end{cases}.
\end{align}
Only the self-energies $\Sigma_{n_1}$ in the transition region are updated,
while $\Sigma_\text{metal}$ and $\Sigma_\text{ins}$ are fixed boundary
conditions taken from the single-site DMFT solution. Our setup assumes the
interface is perpendicular to one of the crystal directions ($x$) and the
system is translationally invariant in the other two ($y$ and $z$) so
self-energies are independent of $n_2$ and $n_3$.

To render the equations soluble in the transition region, we compute the
lattice Green's function and use its local component $G_{n_1n_1}$ to map the
system to a chain of $N$ auxiliary impurity problems~\cite{Bakalov2016},
\begin{equation}
  G_{n_1 n_1}(\iwn)=\frac{1}{\iwn-E_\text{imp} -\Delta_{n_1}(\iwn)-\Sigma_{n_1}(\iwn)}.
\end{equation}
Using the extracted impurity levels $E_\text{imp}$ and hybridization functions
$\Delta_{n_1}$, we obtain the new local self-energies $\Sigma_{n_1}$ and
iterate to convergence. The procedure for computing the local Green's function
is provided in the Supplementary Material.





\begin{figure}
  \includegraphics[width=\columnwidth]{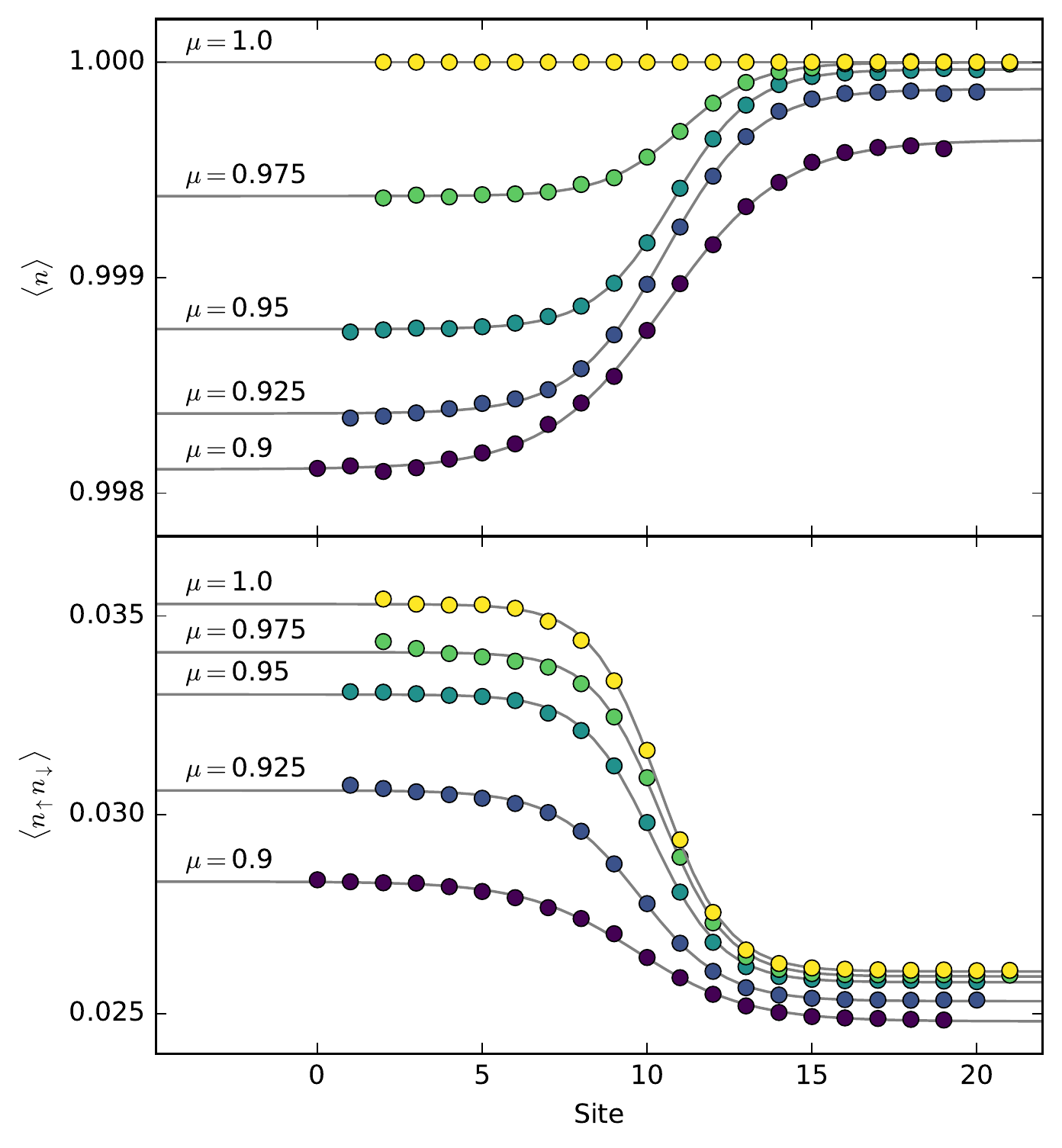}
  \caption{Variation of the density (top) and double occupancy (bottom) across
    the interface at several points along the Mott transition line at $T =
    0.01D$. Thin lines are fits to the standard solution for a double-well
    potential $a + b \tanh((x_i - x_0)/2l)$, allowing extraction of the
    parameters for the underlying free energy. Curves are shifted horizontally
    by varying amounts for clarity. The chemical potential is in units of
    $U/2$, as detailed in the right-hand table of Table~\ref{tbl:lg-params}}
  \label{fig:wall}
\end{figure}

\section{Results}

The evolution of the density $n$ double occupancy $d$ for along the Mott
transition line is displayed in Fig.~\ref{fig:wall} at a temperature of $T =
0.005D$. At the particle-hole symmetric point, there is no jump in density
between the metal and Mott insulator, while the change in double-occupancy is
maximal. As we progress along the transition line (which we parameterize with
the chemical potential $\mu$) towards the hole-doped side, the density
difference between the metal and insulator increases. Additionally, the density
of the insulator drops below unity because we are at finite temperature. In
contrast, the jump in double occupancy decreases.

\begin{figure}
  \includegraphics[width=\columnwidth]{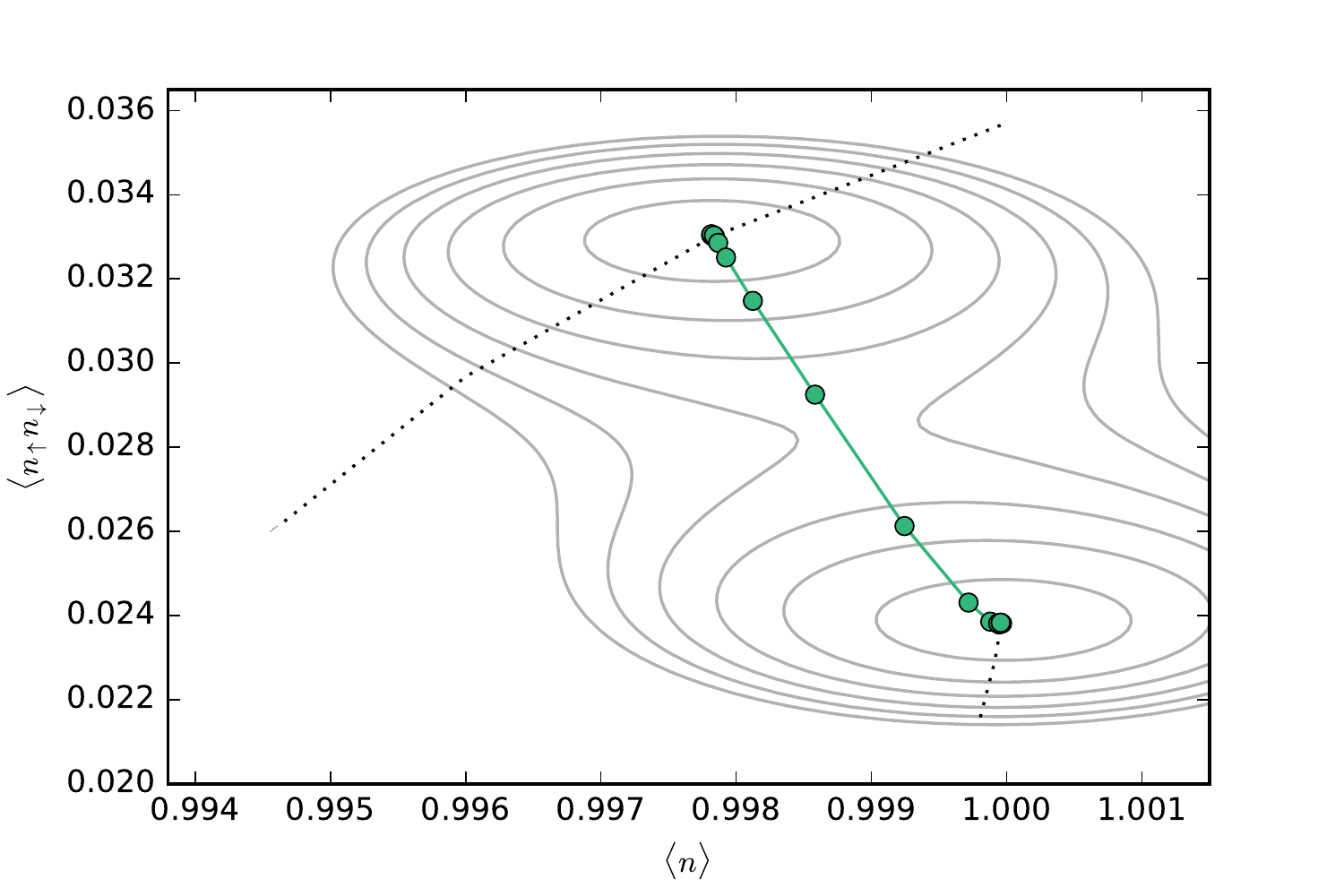}
  \caption{Trajectory in $(n,d)$ space as the system evolves across the
    interface from the insulating to the metallic minima at $\mu = 0.95(U/2)$,
    $U = 2.05D$, $T = 0.005D$. The extracted Landau parameters are used to plot
    the contours of the double-well potential. The dotted lines trace the shift
    of the minima along the Mott transition line at $T = 0.005D$.}
  \label{fig:contour}
\end{figure}

The variation of both quantities fit well to Eq.~\ref{eqn:soliton} for the
double well potential, albeit with slightly different length scales, and we use
the average of the two wall widths to compute $\lambda$. The small difference
in length scales implies the potential is not perfectly symmetric (as
expected), and that the path in $(n,d)$ space between the two minima is close
to, but not exactly, a straight line (see Fig.~\ref{fig:contour}). The
extracted parameters for the Landau free energy are presented in
Table~\ref{tbl:lg-params}.

\begin{table*}
  \centering
  \begin{tabular}{cc|ccccc|c}
    $\mu/(U/2)$ & $U/D$ & $\ni$ & $\di$ & $\nm$ & $\dm$ & $\lambda/D$ & $\alpha$ \\
    \hline
    \hline
    1.00 & 2.04  &  1.0000 & 0.0241 & 1.0000 & 0.0357 & 2410  &  0$^\circ$ \\
    0.95 & 2.05  &  1.0000 & 0.0238 & 0.9978 & 0.0330 & 3170  & 13$^\circ$ \\
    0.90 & 2.08  &  0.9999 & 0.0229 & 0.9960 & 0.0297 & 2510  & 30$^\circ$ \\
    0.85 & 2.13  &  0.9998 & 0.0216 & 0.9947 & 0.0262 & 1780  & 48$^\circ$ \\
    \hline
  \end{tabular}
  \hspace{10pt}
  \begin{tabular}{cc|ccccc|c}
    $\mu/(U/2)$ & $U/D$ & $\ni$ & $\di$ & $\nm$ & $\dm$ & $\lambda/D$ & $\alpha$ \\
    \hline
    \hline
    1.000 & 1.962  &  1.0000 & 0.0261 & 1.0000 & 0.0353 & 1420  &  0$^\circ$ \\
    0.975 & 1.965  &  1.0000 & 0.0259 & 0.9997 & 0.0341 & 1590  &  2$^\circ$ \\
    0.950 & 1.970  &  1.0000 & 0.0258 & 0.9994 & 0.0330 & 1890  &  5$^\circ$ \\
    0.925 & 1.985  &  0.9999 & 0.0253 & 0.9992 & 0.0306 & 2480  &  8$^\circ$ \\
    0.900 & 2.005  &  0.9998 & 0.0248 & 0.9991 & 0.0283 & 3190  & 12$^\circ$ \\
    \hline
  \end{tabular}
  \caption{Extracted parameters of Landau free energy for $T = 0.005D$ (left)
    and $T = 0.01D$ (right), where $D = 6t$ is the half-bandwidth. The position
    along the Mott transition line is parameterized by the chemical potential
    $\mu$, or equivalently, the electron repulsion. The shifts in the density
    and double-occupancy for the Mott insulator $(\ni,\di)$ and metal
    $(\nm,\dm)$ are quite small for the one-band model, which when combined
    with fact that the interface widths $l \sim O(1)$, produces large values of
    $\lambda/D$. The angle $\alpha$ specifies how much of $n$ is admixed into
    the $d$ to form the Ising order parameter (see Eq.~\ref{eqn:op}).}
  \label{tbl:lg-params}
\end{table*}

We also analytically continue the Matsubara self-energies produced by the
impurity solver to the real axis to compute the variation of the spectral
density across the interface. We plot in Fig.~\ref{fig:spectra} the spectra for
parameters $\mu = 0.95(U/2)$, $U = 1.97D$ and $T = 0.01D$, which is slightly on
the hole-doped side. Starting from the metallic solution, we find that the
quasiparticle peak shift slightly downwards and disappears into the lower
Hubbard band as we progress to the Mott insulator. The gap between the Hubbard
bands slightly narrow.

The extracted parameters combined with our ansatz (Eq.~\ref{eqn:free}) allow us
to reconstruct the free energy. Shown in Fig.~\ref{fig:contour} is a
representative case for $\mu = 0.95$, $U = 2.05$, $T = 0.005D$. We have plotted
the trajectory in $(n,d)$ space as the system evolves from the metallic to
insulating minima, superimposed with contour lines of the potential constructed
using the extracted parameters. The movement of the two minima as we step along
the Mott transition line is shown in the dotted lines.

As promised, we explicitly construct the order parameter field $\Delta$ as a
linear combination of $n$ and $d$, owing to the fact that the trajectory is
almost straight. The construction is essentially geometric: we take the line
segment joining the two minima and parameterize it with an angle $\alpha$:
\begin{equation}
  \Delta = (n - \bar{n}) \sin\alpha + (d - \bar{d}) \cos\alpha
  \label{eqn:op}
\end{equation}
where $\bar{n} = (n_i + n_m)/2$ and $\bar{d} = (d_i + d_m)/2$. The angles are
tabulated in Table~\ref{tbl:lg-params}. At particle-hole symmetry, the angle is
zero and the variation of the order parameter is entirely driven by the double
occupancy. Increasing angles imply the density becomes a larger component of
the order parameter, which occurs as we progress to larger correlation
strengths.


\section{Summary}

In this work, we have taken a first step towards characterizing the metal-Mott
interface by modeling its spatial properties, constructing a Landau free energy
and identifying an Ising order parameter. The key parameter of the free energy
which could not be obtained by previous solutions in homogenous geometries is
the interface width $l$, which is directly related to the double-well barrier
height via $\lambda/D$. We also comment that while in general for first-order
transitions which do not possess an organizing symmetry, any number of fields
can be chosen to construct the free energy~\cite{Kotliar2000, Onoda2003,
  Vucicevic2013}, the choice of the quantities conjugate to the physical tuning
parameters $\mu$ and $U$ allow for an especially transparent construction of
the order parameter which can uniformly treat both the bandwidth and
filling controlled transitions.


We want to point out the simplifying assumptions used: (1) we took the
interface to be perpendicular to a crystallographic axis, (2) we only included
nearest-neighbor hopping to simplify the formulae, (3) we made the slow-varying
approximation, assuming each site was an independent impurity affecting the
others only via the hybridization, and (4) we have ignored the long-range
Coulomb interaction. Relaxing these assumptions to capture more realistic
scenarios warrant further investigation.

We expect that future calculations on realistic systems will provide
quantitative results for comparision with near-field optics and STM
observations, and more speculatively, could provide a new constraint on the
value of $U$ in these compounds. Finally, while we have guessed the form of the
Landau free energy and numerically determined its parameters, especially
satisfying for future work would be a microscopic derivation from the
appropriate mean-field theory.

\section{Acknowledgements}

C.Y. thanks Patrick Semon, Camille Aron, Premala Chandra, and Gabriel Kotliar
for stimulating discussions, and Leon Balents, whose thermodynamic construction
of the Mott transition inspired this work. C.Y. was supported as part of the
Center for Emergent Superconductivity, an Energy Frontier Research Center
funded by the US Department of Energy, Office of Science, Office of Basic
Energy Sciences under Award No.~DEAC0298CH1088. J.L. acknowledges support from
the Rutgers Physics Departmental Fellowship.

\bibliography{metal-mott-interface}

\begin{thebibliography}{36}%
\makeatletter
\providecommand \@ifxundefined [1]{%
 \@ifx{#1\undefined}
}%
\providecommand \@ifnum [1]{%
 \ifnum #1\expandafter \@firstoftwo
 \else \expandafter \@secondoftwo
 \fi
}%
\providecommand \@ifx [1]{%
 \ifx #1\expandafter \@firstoftwo
 \else \expandafter \@secondoftwo
 \fi
}%
\providecommand \natexlab [1]{#1}%
\providecommand \enquote  [1]{``#1''}%
\providecommand \bibnamefont  [1]{#1}%
\providecommand \bibfnamefont [1]{#1}%
\providecommand \citenamefont [1]{#1}%
\providecommand \href@noop [0]{\@secondoftwo}%
\providecommand \href [0]{\begingroup \@sanitize@url \@href}%
\providecommand \@href[1]{\@@startlink{#1}\@@href}%
\providecommand \@@href[1]{\endgroup#1\@@endlink}%
\providecommand \@sanitize@url [0]{\catcode `\\12\catcode `\$12\catcode
  `\&12\catcode `\#12\catcode `\^12\catcode `\_12\catcode `\%12\relax}%
\providecommand \@@startlink[1]{}%
\providecommand \@@endlink[0]{}%
\providecommand \url  [0]{\begingroup\@sanitize@url \@url }%
\providecommand \@url [1]{\endgroup\@href {#1}{\urlprefix }}%
\providecommand \urlprefix  [0]{URL }%
\providecommand \Eprint [0]{\href }%
\providecommand \doibase [0]{http://dx.doi.org/}%
\providecommand \selectlanguage [0]{\@gobble}%
\providecommand \bibinfo  [0]{\@secondoftwo}%
\providecommand \bibfield  [0]{\@secondoftwo}%
\providecommand \translation [1]{[#1]}%
\providecommand \BibitemOpen [0]{}%
\providecommand \bibitemStop [0]{}%
\providecommand \bibitemNoStop [0]{.\EOS\space}%
\providecommand \EOS [0]{\spacefactor3000\relax}%
\providecommand \BibitemShut  [1]{\csname bibitem#1\endcsname}%
\let\auto@bib@innerbib\@empty
\bibitem [{\citenamefont {Binder}(1987)}]{Binder1987}%
  \BibitemOpen
  \bibfield  {author} {\bibinfo {author} {\bibfnamefont {K.}~\bibnamefont
  {Binder}},\ }\href {\doibase 10.1088/0034-4885/50/7/001} {\bibfield
  {journal} {\bibinfo  {journal} {Reports on Progress in Physics}\ }\textbf
  {\bibinfo {volume} {50}},\ \bibinfo {pages} {783} (\bibinfo {year}
  {1987})}\BibitemShut {NoStop}%
\bibitem [{\citenamefont {Imada}\ \emph {et~al.}(1998)\citenamefont {Imada},
  \citenamefont {Fujimori},\ and\ \citenamefont {Tokura}}]{Imada1998}%
  \BibitemOpen
  \bibfield  {author} {\bibinfo {author} {\bibfnamefont {M.}~\bibnamefont
  {Imada}}, \bibinfo {author} {\bibfnamefont {A.}~\bibnamefont {Fujimori}}, \
  and\ \bibinfo {author} {\bibfnamefont {Y.}~\bibnamefont {Tokura}},\ }\href
  {\doibase 10.1103/RevModPhys.70.1039} {\bibfield  {journal} {\bibinfo
  {journal} {Reviews of Modern Physics}\ }\textbf {\bibinfo {volume} {70}},\
  \bibinfo {pages} {1039} (\bibinfo {year} {1998})}\BibitemShut {NoStop}%
\bibitem [{\citenamefont {Visscher}(1974)}]{Visscher1974}%
  \BibitemOpen
  \bibfield  {author} {\bibinfo {author} {\bibfnamefont {P.}~\bibnamefont
  {Visscher}},\ }\href {\doibase 10.1103/PhysRevB.10.943} {\bibfield  {journal}
  {\bibinfo  {journal} {Physical Review B}\ }\textbf {\bibinfo {volume} {10}},\
  \bibinfo {pages} {943} (\bibinfo {year} {1974})}\BibitemShut {NoStop}%
\bibitem [{\citenamefont {Emery}\ \emph {et~al.}(1990)\citenamefont {Emery},
  \citenamefont {Kivelson},\ and\ \citenamefont {Lin}}]{Emery1990}%
  \BibitemOpen
  \bibfield  {author} {\bibinfo {author} {\bibfnamefont {V.}~\bibnamefont
  {Emery}}, \bibinfo {author} {\bibfnamefont {S.}~\bibnamefont {Kivelson}}, \
  and\ \bibinfo {author} {\bibfnamefont {H.}~\bibnamefont {Lin}},\ }\href
  {\doibase 10.1103/PhysRevLett.64.475} {\bibfield  {journal} {\bibinfo
  {journal} {Physical Review Letters}\ }\textbf {\bibinfo {volume} {64}},\
  \bibinfo {pages} {475} (\bibinfo {year} {1990})}\BibitemShut {NoStop}%
\bibitem [{\citenamefont {Gehlhoff}(1996)}]{Gehlhoff1996}%
  \BibitemOpen
  \bibfield  {author} {\bibinfo {author} {\bibfnamefont {L.}~\bibnamefont
  {Gehlhoff}},\ }\href {\doibase 10.1088/0953-8984/8/16/014} {\bibfield
  {journal} {\bibinfo  {journal} {Journal of Physics: Condensed Matter}\
  }\textbf {\bibinfo {volume} {8}},\ \bibinfo {pages} {2851} (\bibinfo {year}
  {1996})}\BibitemShut {NoStop}%
\bibitem [{\citenamefont {White}\ and\ \citenamefont
  {Scalapino}(2000)}]{White2000}%
  \BibitemOpen
  \bibfield  {author} {\bibinfo {author} {\bibfnamefont {S.}~\bibnamefont
  {White}}\ and\ \bibinfo {author} {\bibfnamefont {D.}~\bibnamefont
  {Scalapino}},\ }\href {\doibase 10.1103/PhysRevB.61.6320} {\bibfield
  {journal} {\bibinfo  {journal} {Physical Review B}\ }\textbf {\bibinfo
  {volume} {61}},\ \bibinfo {pages} {6320} (\bibinfo {year}
  {2000})}\BibitemShut {NoStop}%
\bibitem [{\citenamefont {Galanakis}\ \emph {et~al.}(2011)\citenamefont
  {Galanakis}, \citenamefont {Khatami}, \citenamefont {Mikelsons},
  \citenamefont {Macridin}, \citenamefont {Moreno}, \citenamefont {Browne},\
  and\ \citenamefont {Jarrell}}]{Galanakis2011}%
  \BibitemOpen
  \bibfield  {author} {\bibinfo {author} {\bibfnamefont {D.}~\bibnamefont
  {Galanakis}}, \bibinfo {author} {\bibfnamefont {E.}~\bibnamefont {Khatami}},
  \bibinfo {author} {\bibfnamefont {K.}~\bibnamefont {Mikelsons}}, \bibinfo
  {author} {\bibfnamefont {a.}~\bibnamefont {Macridin}}, \bibinfo {author}
  {\bibfnamefont {J.}~\bibnamefont {Moreno}}, \bibinfo {author} {\bibfnamefont
  {D.~a.}\ \bibnamefont {Browne}}, \ and\ \bibinfo {author} {\bibfnamefont
  {M.}~\bibnamefont {Jarrell}},\ }\href {\doibase 10.1098/rsta.2010.0228}
  {\bibfield  {journal} {\bibinfo  {journal} {Philosophical transactions.
  Series A, Mathematical, physical, and engineering sciences}\ }\textbf
  {\bibinfo {volume} {369}},\ \bibinfo {pages} {1670} (\bibinfo {year}
  {2011})}\BibitemShut {NoStop}%
\bibitem [{\citenamefont {Yee}\ and\ \citenamefont {Balents}(2015)}]{Yee2015a}%
  \BibitemOpen
  \bibfield  {author} {\bibinfo {author} {\bibfnamefont {C.-h.}\ \bibnamefont
  {Yee}}\ and\ \bibinfo {author} {\bibfnamefont {L.}~\bibnamefont {Balents}},\
  }\href {\doibase 10.1103/PhysRevX.5.021007} {\bibfield  {journal} {\bibinfo
  {journal} {Physical Review X}\ }\textbf {\bibinfo {volume} {5}},\ \bibinfo
  {pages} {021007} (\bibinfo {year} {2015})},\ \Eprint
  {http://arxiv.org/abs/1407.0368} {arXiv:1407.0368} \BibitemShut {NoStop}%
\bibitem [{\citenamefont {Hanaguri}\ \emph {et~al.}(2004)\citenamefont
  {Hanaguri}, \citenamefont {Kohsaka}, \citenamefont {Iwaya}, \citenamefont
  {Satow}, \citenamefont {Kitazawa}, \citenamefont {Takagi}, \citenamefont
  {Azuma},\ and\ \citenamefont {Takano}}]{Hanaguri2004}%
  \BibitemOpen
  \bibfield  {author} {\bibinfo {author} {\bibfnamefont {T.}~\bibnamefont
  {Hanaguri}}, \bibinfo {author} {\bibfnamefont {Y.}~\bibnamefont {Kohsaka}},
  \bibinfo {author} {\bibfnamefont {K.}~\bibnamefont {Iwaya}}, \bibinfo
  {author} {\bibfnamefont {S.}~\bibnamefont {Satow}}, \bibinfo {author}
  {\bibfnamefont {K.}~\bibnamefont {Kitazawa}}, \bibinfo {author}
  {\bibfnamefont {H.}~\bibnamefont {Takagi}}, \bibinfo {author} {\bibfnamefont
  {M.}~\bibnamefont {Azuma}}, \ and\ \bibinfo {author} {\bibfnamefont
  {M.}~\bibnamefont {Takano}},\ }\href {\doibase 10.1016/j.physc.2004.02.133}
  {\bibfield  {journal} {\bibinfo  {journal} {Physica C: Superconductivity and
  its Applications}\ }\textbf {\bibinfo {volume} {408-410}},\ \bibinfo {pages}
  {328} (\bibinfo {year} {2004})}\BibitemShut {NoStop}%
\bibitem [{\citenamefont {Qazilbash}\ \emph {et~al.}(2007)\citenamefont
  {Qazilbash}, \citenamefont {Brehm}, \citenamefont {Chae}, \citenamefont {Ho},
  \citenamefont {Andreev}, \citenamefont {Kim}, \citenamefont {Yun},
  \citenamefont {Balatsky}, \citenamefont {Maple}, \citenamefont {Keilmann},
  \citenamefont {Kim},\ and\ \citenamefont {Basov}}]{Qazilbash2007}%
  \BibitemOpen
  \bibfield  {author} {\bibinfo {author} {\bibfnamefont {M.~M.}\ \bibnamefont
  {Qazilbash}}, \bibinfo {author} {\bibfnamefont {M.}~\bibnamefont {Brehm}},
  \bibinfo {author} {\bibfnamefont {B.-G.}\ \bibnamefont {Chae}}, \bibinfo
  {author} {\bibfnamefont {P.-C.}\ \bibnamefont {Ho}}, \bibinfo {author}
  {\bibfnamefont {G.~O.}\ \bibnamefont {Andreev}}, \bibinfo {author}
  {\bibfnamefont {B.-J.}\ \bibnamefont {Kim}}, \bibinfo {author} {\bibfnamefont
  {S.~J.}\ \bibnamefont {Yun}}, \bibinfo {author} {\bibfnamefont {A.~V.}\
  \bibnamefont {Balatsky}}, \bibinfo {author} {\bibfnamefont {M.~B.}\
  \bibnamefont {Maple}}, \bibinfo {author} {\bibfnamefont {F.}~\bibnamefont
  {Keilmann}}, \bibinfo {author} {\bibfnamefont {H.-T.}\ \bibnamefont {Kim}}, \
  and\ \bibinfo {author} {\bibfnamefont {D.~N.}\ \bibnamefont {Basov}},\ }\href
  {\doibase 10.1126/science.1150124} {\bibfield  {journal} {\bibinfo  {journal}
  {Science (New York, N.Y.)}\ }\textbf {\bibinfo {volume} {318}},\ \bibinfo
  {pages} {1750} (\bibinfo {year} {2007})}\BibitemShut {NoStop}%
\bibitem [{\citenamefont {Kohsaka}\ \emph {et~al.}(2012)\citenamefont
  {Kohsaka}, \citenamefont {Hanaguri}, \citenamefont {Azuma}, \citenamefont
  {Takano}, \citenamefont {Davis},\ and\ \citenamefont {Takagi}}]{Kohsaka2012}%
  \BibitemOpen
  \bibfield  {author} {\bibinfo {author} {\bibfnamefont {Y.}~\bibnamefont
  {Kohsaka}}, \bibinfo {author} {\bibfnamefont {T.}~\bibnamefont {Hanaguri}},
  \bibinfo {author} {\bibfnamefont {M.}~\bibnamefont {Azuma}}, \bibinfo
  {author} {\bibfnamefont {M.}~\bibnamefont {Takano}}, \bibinfo {author}
  {\bibfnamefont {J.~C.}\ \bibnamefont {Davis}}, \ and\ \bibinfo {author}
  {\bibfnamefont {H.}~\bibnamefont {Takagi}},\ }\href {\doibase
  10.1038/nphys2321} {\bibfield  {journal} {\bibinfo  {journal} {Nature
  Physics}\ }\textbf {\bibinfo {volume} {8}},\ \bibinfo {pages} {1} (\bibinfo
  {year} {2012})}\BibitemShut {NoStop}%
\bibitem [{\citenamefont {Schwieger}\ \emph {et~al.}(2003)\citenamefont
  {Schwieger}, \citenamefont {Potthoff},\ and\ \citenamefont
  {Nolting}}]{Schwieger2003}%
  \BibitemOpen
  \bibfield  {author} {\bibinfo {author} {\bibfnamefont {S.}~\bibnamefont
  {Schwieger}}, \bibinfo {author} {\bibfnamefont {M.}~\bibnamefont {Potthoff}},
  \ and\ \bibinfo {author} {\bibfnamefont {W.}~\bibnamefont {Nolting}},\ }\href
  {\doibase 10.1103/PhysRevB.67.165408} {\bibfield  {journal} {\bibinfo
  {journal} {Physical Review B}\ }\textbf {\bibinfo {volume} {67}},\ \bibinfo
  {pages} {8} (\bibinfo {year} {2003})},\ \Eprint
  {http://arxiv.org/abs/0302427} {arXiv:0302427 [cond-mat]} \BibitemShut
  {NoStop}%
\bibitem [{\citenamefont {Ishida}\ \emph {et~al.}(2006)\citenamefont {Ishida},
  \citenamefont {Wortmann},\ and\ \citenamefont {Liebsch}}]{Ishida2006}%
  \BibitemOpen
  \bibfield  {author} {\bibinfo {author} {\bibfnamefont {H.}~\bibnamefont
  {Ishida}}, \bibinfo {author} {\bibfnamefont {D.}~\bibnamefont {Wortmann}}, \
  and\ \bibinfo {author} {\bibfnamefont {A.}~\bibnamefont {Liebsch}},\ }\href
  {\doibase 10.1103/PhysRevB.73.245421} {\bibfield  {journal} {\bibinfo
  {journal} {Physical Review B}\ }\textbf {\bibinfo {volume} {73}},\ \bibinfo
  {pages} {245421} (\bibinfo {year} {2006})}\BibitemShut {NoStop}%
\bibitem [{\citenamefont {Freericks}(2004)}]{Freericks2004}%
  \BibitemOpen
  \bibfield  {author} {\bibinfo {author} {\bibfnamefont {J.~K.}\ \bibnamefont
  {Freericks}},\ }\href {\doibase 10.1103/PhysRevB.70.195342} {\bibfield
  {journal} {\bibinfo  {journal} {Physical Review B - Condensed Matter and
  Materials Physics}\ }\textbf {\bibinfo {volume} {70}},\ \bibinfo {pages} {1}
  (\bibinfo {year} {2004})},\ \Eprint {http://arxiv.org/abs/0408226}
  {arXiv:0408226 [cond-mat]} \BibitemShut {NoStop}%
\bibitem [{\citenamefont {Helmes}\ \emph {et~al.}(2008)\citenamefont {Helmes},
  \citenamefont {Costi},\ and\ \citenamefont {Rosch}}]{Helmes2008}%
  \BibitemOpen
  \bibfield  {author} {\bibinfo {author} {\bibfnamefont {R.}~\bibnamefont
  {Helmes}}, \bibinfo {author} {\bibfnamefont {T.}~\bibnamefont {Costi}}, \
  and\ \bibinfo {author} {\bibfnamefont {a.}~\bibnamefont {Rosch}},\ }\href
  {\doibase 10.1103/PhysRevLett.101.066802} {\bibfield  {journal} {\bibinfo
  {journal} {Physical Review Letters}\ }\textbf {\bibinfo {volume} {101}},\
  \bibinfo {pages} {066802} (\bibinfo {year} {2008})}\BibitemShut {NoStop}%
\bibitem [{\citenamefont {Zenia}\ \emph {et~al.}(2009)\citenamefont {Zenia},
  \citenamefont {Freericks}, \citenamefont {Krishnamurthy},\ and\ \citenamefont
  {Pruschke}}]{Zenia2009}%
  \BibitemOpen
  \bibfield  {author} {\bibinfo {author} {\bibfnamefont {H.}~\bibnamefont
  {Zenia}}, \bibinfo {author} {\bibfnamefont {J.~K.}\ \bibnamefont
  {Freericks}}, \bibinfo {author} {\bibfnamefont {H.~R.}\ \bibnamefont
  {Krishnamurthy}}, \ and\ \bibinfo {author} {\bibfnamefont {T.}~\bibnamefont
  {Pruschke}},\ }\href {\doibase 10.1103/PhysRevLett.103.116402} {\bibfield
  {journal} {\bibinfo  {journal} {Physical Review Letters}\ }\textbf {\bibinfo
  {volume} {103}},\ \bibinfo {pages} {1} (\bibinfo {year} {2009})}\BibitemShut
  {NoStop}%
\bibitem [{\citenamefont {Borghi}\ \emph {et~al.}(2010)\citenamefont {Borghi},
  \citenamefont {Fabrizio},\ and\ \citenamefont {Tosatti}}]{Borghi2010}%
  \BibitemOpen
  \bibfield  {author} {\bibinfo {author} {\bibfnamefont {G.}~\bibnamefont
  {Borghi}}, \bibinfo {author} {\bibfnamefont {M.}~\bibnamefont {Fabrizio}}, \
  and\ \bibinfo {author} {\bibfnamefont {E.}~\bibnamefont {Tosatti}},\ }\href
  {\doibase 10.1103/PhysRevB.81.115134} {\bibfield  {journal} {\bibinfo
  {journal} {Physical Review B - Condensed Matter and Materials Physics}\
  }\textbf {\bibinfo {volume} {81}},\ \bibinfo {pages} {1} (\bibinfo {year}
  {2010})},\ \Eprint {http://arxiv.org/abs/0911.0718} {arXiv:0911.0718}
  \BibitemShut {NoStop}%
\bibitem [{\citenamefont {Bakalov}\ \emph {et~al.}(2016)\citenamefont
  {Bakalov}, \citenamefont {{Nasr Esfahani}}, \citenamefont {Covaci},
  \citenamefont {Peeters}, \citenamefont {Tempere},\ and\ \citenamefont
  {Locquet}}]{Bakalov2016}%
  \BibitemOpen
  \bibfield  {author} {\bibinfo {author} {\bibfnamefont {P.}~\bibnamefont
  {Bakalov}}, \bibinfo {author} {\bibfnamefont {D.}~\bibnamefont {{Nasr
  Esfahani}}}, \bibinfo {author} {\bibfnamefont {L.}~\bibnamefont {Covaci}},
  \bibinfo {author} {\bibfnamefont {F.~M.}\ \bibnamefont {Peeters}}, \bibinfo
  {author} {\bibfnamefont {J.}~\bibnamefont {Tempere}}, \ and\ \bibinfo
  {author} {\bibfnamefont {J.~P.}\ \bibnamefont {Locquet}},\ }\href {\doibase
  10.1103/PhysRevB.93.165112} {\bibfield  {journal} {\bibinfo  {journal}
  {Physical Review B - Condensed Matter and Materials Physics}\ }\textbf
  {\bibinfo {volume} {93}},\ \bibinfo {pages} {1} (\bibinfo {year} {2016})},\
  \Eprint {http://arxiv.org/abs/1503.02502} {arXiv:1503.02502} \BibitemShut
  {NoStop}%
\bibitem [{\citenamefont {Castellani}\ \emph {et~al.}(1979)\citenamefont
  {Castellani}, \citenamefont {{Di Castro}}, \citenamefont {Feinberg},\ and\
  \citenamefont {Ranninger}}]{Castellani1979}%
  \BibitemOpen
  \bibfield  {author} {\bibinfo {author} {\bibfnamefont {C.}~\bibnamefont
  {Castellani}}, \bibinfo {author} {\bibfnamefont {C.}~\bibnamefont {{Di
  Castro}}}, \bibinfo {author} {\bibfnamefont {D.}~\bibnamefont {Feinberg}}, \
  and\ \bibinfo {author} {\bibfnamefont {J.}~\bibnamefont {Ranninger}},\ }\href
  {\doibase 10.1103/PhysRevLett.43.1957} {\bibfield  {journal} {\bibinfo
  {journal} {Physical Review Letters}\ }\textbf {\bibinfo {volume} {43}},\
  \bibinfo {pages} {1957} (\bibinfo {year} {1979})}\BibitemShut {NoStop}%
\bibitem [{\citenamefont {Kotliar}\ \emph {et~al.}(2000)\citenamefont
  {Kotliar}, \citenamefont {Lange},\ and\ \citenamefont
  {Rozenberg}}]{Kotliar2000}%
  \BibitemOpen
  \bibfield  {author} {\bibinfo {author} {\bibfnamefont {G.}~\bibnamefont
  {Kotliar}}, \bibinfo {author} {\bibfnamefont {E.}~\bibnamefont {Lange}}, \
  and\ \bibinfo {author} {\bibfnamefont {M.}~\bibnamefont {Rozenberg}},\ }\href
  {http://www.ncbi.nlm.nih.gov/pubmed/10990897} {\bibfield  {journal} {\bibinfo
   {journal} {Physical review letters}\ }\textbf {\bibinfo {volume} {84}},\
  \bibinfo {pages} {5180} (\bibinfo {year} {2000})}\BibitemShut {NoStop}%
\bibitem [{\citenamefont {Onoda}\ and\ \citenamefont
  {Nagaosa}(2003)}]{Onoda2003}%
  \BibitemOpen
  \bibfield  {author} {\bibinfo {author} {\bibfnamefont {S.}~\bibnamefont
  {Onoda}}\ and\ \bibinfo {author} {\bibfnamefont {N.}~\bibnamefont
  {Nagaosa}},\ }\href {\doibase 10.1143/JPSJ.72.2445} {\bibfield  {journal}
  {\bibinfo  {journal} {Journal of the Physical Society of Japan}\ }\textbf
  {\bibinfo {volume} {72}},\ \bibinfo {pages} {2445} (\bibinfo {year}
  {2003})}\BibitemShut {NoStop}%
\bibitem [{\citenamefont {Limelette}\ \emph {et~al.}(2003)\citenamefont
  {Limelette}, \citenamefont {Georges}, \citenamefont {J{\'{e}}rome},\ and\
  \citenamefont {Wzietek}}]{Limelette2003}%
  \BibitemOpen
  \bibfield  {author} {\bibinfo {author} {\bibfnamefont {P.}~\bibnamefont
  {Limelette}}, \bibinfo {author} {\bibfnamefont {A.}~\bibnamefont {Georges}},
  \bibinfo {author} {\bibfnamefont {D.}~\bibnamefont {J{\'{e}}rome}}, \ and\
  \bibinfo {author} {\bibfnamefont {P.}~\bibnamefont {Wzietek}},\ }\href
  {http://www.sciencemag.org/content/302/5642/89.short} {\bibfield  {journal}
  {\bibinfo  {journal} {Science}\ }\textbf {\bibinfo {volume} {302}},\ \bibinfo
  {pages} {89} (\bibinfo {year} {2003})}\BibitemShut {NoStop}%
\bibitem [{\citenamefont {Kagawa}\ \emph {et~al.}(2005)\citenamefont {Kagawa},
  \citenamefont {Miyagawa},\ and\ \citenamefont {Kanoda}}]{Kagawa2005}%
  \BibitemOpen
  \bibfield  {author} {\bibinfo {author} {\bibfnamefont {F.}~\bibnamefont
  {Kagawa}}, \bibinfo {author} {\bibfnamefont {K.}~\bibnamefont {Miyagawa}}, \
  and\ \bibinfo {author} {\bibfnamefont {K.}~\bibnamefont {Kanoda}},\ }\href
  {\doibase 10.1038/nature03806} {\bibfield  {journal} {\bibinfo  {journal}
  {Nature}\ }\textbf {\bibinfo {volume} {436}},\ \bibinfo {pages} {534}
  (\bibinfo {year} {2005})},\ \Eprint {http://arxiv.org/abs/0603064}
  {arXiv:0603064 [cond-mat]} \BibitemShut {NoStop}%
\bibitem [{\citenamefont {Papanikolaou}\ \emph {et~al.}(2008)\citenamefont
  {Papanikolaou}, \citenamefont {Fernandes}, \citenamefont {Fradkin},
  \citenamefont {Phillips}, \citenamefont {Schmalian},\ and\ \citenamefont
  {Sknepnek}}]{Papanikolaou2008}%
  \BibitemOpen
  \bibfield  {author} {\bibinfo {author} {\bibfnamefont {S.}~\bibnamefont
  {Papanikolaou}}, \bibinfo {author} {\bibfnamefont {R.~M.}\ \bibnamefont
  {Fernandes}}, \bibinfo {author} {\bibfnamefont {E.}~\bibnamefont {Fradkin}},
  \bibinfo {author} {\bibfnamefont {P.~W.}\ \bibnamefont {Phillips}}, \bibinfo
  {author} {\bibfnamefont {J.}~\bibnamefont {Schmalian}}, \ and\ \bibinfo
  {author} {\bibfnamefont {R.}~\bibnamefont {Sknepnek}},\ }\href {\doibase
  10.1103/PhysRevLett.100.026408} {\bibfield  {journal} {\bibinfo  {journal}
  {Physical Review Letters}\ }\textbf {\bibinfo {volume} {100}},\ \bibinfo
  {pages} {1} (\bibinfo {year} {2008})},\ \Eprint
  {http://arxiv.org/abs/0710.1627} {arXiv:0710.1627} \BibitemShut {NoStop}%
\bibitem [{\citenamefont {S{\'{e}}mon}\ and\ \citenamefont
  {Tremblay}(2012)}]{Semon2012}%
  \BibitemOpen
  \bibfield  {author} {\bibinfo {author} {\bibfnamefont {P.}~\bibnamefont
  {S{\'{e}}mon}}\ and\ \bibinfo {author} {\bibfnamefont {a.-M.~S.}\
  \bibnamefont {Tremblay}},\ }\href {\doibase 10.1103/PhysRevB.85.201101}
  {\bibfield  {journal} {\bibinfo  {journal} {Physical Review B}\ }\textbf
  {\bibinfo {volume} {85}},\ \bibinfo {pages} {201101} (\bibinfo {year}
  {2012})}\BibitemShut {NoStop}%
\bibitem [{\citenamefont {Vu{\v{c}}i{\v{c}}evi{\'{c}}}\ \emph
  {et~al.}(2013)\citenamefont {Vu{\v{c}}i{\v{c}}evi{\'{c}}}, \citenamefont
  {Terletska}, \citenamefont {Tanaskovi{\'{c}}},\ and\ \citenamefont
  {Dobrosavljevi{\'{c}}}}]{Vucicevic2013}%
  \BibitemOpen
  \bibfield  {author} {\bibinfo {author} {\bibfnamefont {J.}~\bibnamefont
  {Vu{\v{c}}i{\v{c}}evi{\'{c}}}}, \bibinfo {author} {\bibfnamefont
  {H.}~\bibnamefont {Terletska}}, \bibinfo {author} {\bibfnamefont
  {D.}~\bibnamefont {Tanaskovi{\'{c}}}}, \ and\ \bibinfo {author}
  {\bibfnamefont {V.}~\bibnamefont {Dobrosavljevi{\'{c}}}},\ }\href {\doibase
  10.1103/PhysRevB.88.075143} {\bibfield  {journal} {\bibinfo  {journal}
  {Physical Review B - Condensed Matter and Materials Physics}\ }\textbf
  {\bibinfo {volume} {88}},\ \bibinfo {pages} {1} (\bibinfo {year}
  {2013})}\BibitemShut {NoStop}%
\bibitem [{\citenamefont {Werner}\ and\ \citenamefont
  {Millis}(2007)}]{Werner2007}%
  \BibitemOpen
  \bibfield  {author} {\bibinfo {author} {\bibfnamefont {P.}~\bibnamefont
  {Werner}}\ and\ \bibinfo {author} {\bibfnamefont {A.}~\bibnamefont
  {Millis}},\ }\href {\doibase 10.1103/PhysRevB.75.085108} {\bibfield
  {journal} {\bibinfo  {journal} {Physical Review B}\ }\textbf {\bibinfo
  {volume} {75}},\ \bibinfo {pages} {085108} (\bibinfo {year}
  {2007})}\BibitemShut {NoStop}%
\bibitem [{\citenamefont {Imada}(2005)}]{Imada2005}%
  \BibitemOpen
  \bibfield  {author} {\bibinfo {author} {\bibfnamefont {M.}~\bibnamefont
  {Imada}},\ }\href {\doibase 10.1103/PhysRevB.72.075113} {\bibfield  {journal}
  {\bibinfo  {journal} {Physical Review B - Condensed Matter and Materials
  Physics}\ }\textbf {\bibinfo {volume} {72}},\ \bibinfo {pages} {1} (\bibinfo
  {year} {2005})},\ \Eprint {http://arxiv.org/abs/0506468} {arXiv:0506468
  [cond-mat]} \BibitemShut {NoStop}%
\bibitem [{\citenamefont {McWhan}\ and\ \citenamefont
  {Remeika}(1970)}]{McWhan1970a}%
  \BibitemOpen
  \bibfield  {author} {\bibinfo {author} {\bibfnamefont {D.~B.}\ \bibnamefont
  {McWhan}}\ and\ \bibinfo {author} {\bibfnamefont {J.~P.}\ \bibnamefont
  {Remeika}},\ }\href {\doibase 10.1103/PhysRevB.2.3734} {\bibfield  {journal}
  {\bibinfo  {journal} {Physical Review B}\ }\textbf {\bibinfo {volume} {2}},\
  \bibinfo {pages} {3734} (\bibinfo {year} {1970})}\BibitemShut {NoStop}%
\bibitem [{\citenamefont {Vainshtein}\ \emph {et~al.}(1982)\citenamefont
  {Vainshtein}, \citenamefont {Zakharov}, \citenamefont {Novikov},\ and\
  \citenamefont {Shifman}}]{Vainshtein1982}%
  \BibitemOpen
  \bibfield  {author} {\bibinfo {author} {\bibfnamefont {A.~I.}\ \bibnamefont
  {Vainshtein}}, \bibinfo {author} {\bibfnamefont {V.~I.}\ \bibnamefont
  {Zakharov}}, \bibinfo {author} {\bibfnamefont {V.~A.}\ \bibnamefont
  {Novikov}}, \ and\ \bibinfo {author} {\bibfnamefont {M.~A.}\ \bibnamefont
  {Shifman}},\ }\href {\doibase 10.3367/UFNr.0136.198204a.0553} {\bibfield
  {journal} {\bibinfo  {journal} {Soviet Physics Uspekhi}\ }\textbf {\bibinfo
  {volume} {25}},\ \bibinfo {pages} {195} (\bibinfo {year} {1982})}\BibitemShut
  {NoStop}%
\bibitem [{\citenamefont {Metzner}\ and\ \citenamefont
  {Vollhardt}(1989)}]{Metzner1989}%
  \BibitemOpen
  \bibfield  {author} {\bibinfo {author} {\bibfnamefont {W.}~\bibnamefont
  {Metzner}}\ and\ \bibinfo {author} {\bibfnamefont {D.}~\bibnamefont
  {Vollhardt}},\ }\href {\doibase 10.1103/PhysRevLett.62.324} {\bibfield
  {journal} {\bibinfo  {journal} {Physical Review Letters}\ }\textbf {\bibinfo
  {volume} {62}},\ \bibinfo {pages} {324} (\bibinfo {year} {1989})},\ \Eprint
  {http://arxiv.org/abs/arXiv:1011.1669v3} {arXiv:arXiv:1011.1669v3}
  \BibitemShut {NoStop}%
\bibitem [{\citenamefont {Georges}\ and\ \citenamefont
  {Kotliar}(1992)}]{Georges1992}%
  \BibitemOpen
  \bibfield  {author} {\bibinfo {author} {\bibfnamefont {A.}~\bibnamefont
  {Georges}}\ and\ \bibinfo {author} {\bibfnamefont {G.}~\bibnamefont
  {Kotliar}},\ }\href {\doibase 10.1103/PhysRevB.45.6479} {\bibfield  {journal}
  {\bibinfo  {journal} {Physical Review B}\ }\textbf {\bibinfo {volume} {45}},\
  \bibinfo {pages} {6479} (\bibinfo {year} {1992})}\BibitemShut {NoStop}%
\bibitem [{\citenamefont {Georges}\ \emph {et~al.}(1996)\citenamefont
  {Georges}, \citenamefont {Kotliar}, \citenamefont {Krauth},\ and\
  \citenamefont {Rozenberg}}]{Georges1996}%
  \BibitemOpen
  \bibfield  {author} {\bibinfo {author} {\bibfnamefont {A.}~\bibnamefont
  {Georges}}, \bibinfo {author} {\bibfnamefont {G.}~\bibnamefont {Kotliar}},
  \bibinfo {author} {\bibfnamefont {W.}~\bibnamefont {Krauth}}, \ and\ \bibinfo
  {author} {\bibfnamefont {M.~J.}\ \bibnamefont {Rozenberg}},\ }\href {\doibase
  10.1103/RevModPhys.68.13} {\bibfield  {journal} {\bibinfo  {journal} {Reviews
  of Modern Physics}\ }\textbf {\bibinfo {volume} {68}},\ \bibinfo {pages} {13}
  (\bibinfo {year} {1996})}\BibitemShut {NoStop}%
\bibitem [{\citenamefont {Werner}\ \emph {et~al.}(2006)\citenamefont {Werner},
  \citenamefont {Comanac}, \citenamefont {de' Medici}, \citenamefont {Troyer},\
  and\ \citenamefont {Millis}}]{Werner2006}%
  \BibitemOpen
  \bibfield  {author} {\bibinfo {author} {\bibfnamefont {P.}~\bibnamefont
  {Werner}}, \bibinfo {author} {\bibfnamefont {A.}~\bibnamefont {Comanac}},
  \bibinfo {author} {\bibfnamefont {L.}~\bibnamefont {de' Medici}}, \bibinfo
  {author} {\bibfnamefont {M.}~\bibnamefont {Troyer}}, \ and\ \bibinfo {author}
  {\bibfnamefont {A.}~\bibnamefont {Millis}},\ }\href {\doibase
  10.1103/PhysRevLett.97.076405} {\bibfield  {journal} {\bibinfo  {journal}
  {Physical Review Letters}\ }\textbf {\bibinfo {volume} {97}},\ \bibinfo
  {pages} {076405} (\bibinfo {year} {2006})}\BibitemShut {NoStop}%
\bibitem [{\citenamefont {Haule}(2007)}]{Haule2007b}%
  \BibitemOpen
  \bibfield  {author} {\bibinfo {author} {\bibfnamefont {K.}~\bibnamefont
  {Haule}},\ }\href {\doibase 10.1103/PhysRevB.75.155113} {\bibfield  {journal}
  {\bibinfo  {journal} {Physical Review B}\ }\textbf {\bibinfo {volume} {75}},\
  \bibinfo {pages} {155113} (\bibinfo {year} {2007})}\BibitemShut {NoStop}%
\bibitem [{\citenamefont {Gull}\ \emph {et~al.}(2011)\citenamefont {Gull},
  \citenamefont {Millis}, \citenamefont {Lichtenstein}, \citenamefont
  {Rubtsov}, \citenamefont {Troyer},\ and\ \citenamefont {Werner}}]{Gull2011}%
  \BibitemOpen
  \bibfield  {author} {\bibinfo {author} {\bibfnamefont {E.}~\bibnamefont
  {Gull}}, \bibinfo {author} {\bibfnamefont {A.~J.}\ \bibnamefont {Millis}},
  \bibinfo {author} {\bibfnamefont {A.~I.}\ \bibnamefont {Lichtenstein}},
  \bibinfo {author} {\bibfnamefont {A.~N.}\ \bibnamefont {Rubtsov}}, \bibinfo
  {author} {\bibfnamefont {M.}~\bibnamefont {Troyer}}, \ and\ \bibinfo {author}
  {\bibfnamefont {P.}~\bibnamefont {Werner}},\ }\href {\doibase
  10.1103/RevModPhys.83.349} {\bibfield  {journal} {\bibinfo  {journal}
  {Reviews of Modern Physics}\ }\textbf {\bibinfo {volume} {83}},\ \bibinfo
  {pages} {349} (\bibinfo {year} {2011})}\BibitemShut {NoStop}%
\end{thebibliography}%

\newpage

\section{Supplementary: Calculation of local Green's function}

The Green's function of the lattice is given by
\begin{align}\label{G_sp}
  G_{\mathbf{R}\mathbf{R}'}=[(\iw+\mu) \delta_{\mathbf{R}\mathbf{R}'} -  t_{\mathbf{R}\mathbf{R}'}-\Sigma_{\mathbf{R}\mathbf{R}'}]^{-1}
\end{align}
where $\mathbf{R}$ is a lattice vector $\mathbf{R}=(n_1,n_2,n_3)$ with the
cubic primitive lattice vector and $t_{\mathbf{R}\mathbf{R}'}$ denotes the
nearest neighbor hopping. To see the spatial variation across the two different
phases, we divide the lattice into three regions: metallic ($\cM: -\infty < n_1
\leq 0 $), insulating ($\cI: N+1\leq n_1 <\infty$) and transition ($\cT: 1 \leq
n_1 \leq N$) region. So $\cT$ is sandwiched by $\cM$ and $\cI$. Then we assign
to each site the localized self-energy
$\Sigma_{n_1n_1'}=\delta_{n_1n_1'}\Sigma_{n_1}$ with
\begin{align}
  \Sigma_{n_1}=\left\{\begin{matrix}
  \Sigma_{\mathrm{metal}} & (n_1\in \cM) \\ 
  \Sigma_{n_1} & (n_1\in \cT)\\
  \Sigma_{\mathrm{ins}}& (n_1\in \cI)\\ 
  \end{matrix}\right.
\end{align}
Note that in the metallic and insulating regimes, the self-energy is fixed to
$\Sigma_{\mathrm{metal}}$ and $\Sigma_{\mathrm{ins}}$ respectively, while we
allow the local self-energy in the transition regime to vary across the sites.

The Fourier transformation of Eq. \eqref{G_sp} along $y$ and $z$ directions
gives the following matrix form of Green's function in the mixed representation
$(n_1;k_y,k_z)$ ($n_1$ is the site index of $x$):
\begin{align}\label{G_mit}
  \big[ G(k_y,k_z;\iw)\big]_{n_1n_1'} &= \bigg[\big[(\iw + \mu - \varepsilon(k_y,k_z)\nonumber\\
      &- \Sigma_{n_1}(\iw))\hat I - \hat t~ \big]^{-1}\bigg]_{n_1 n_1'}
\end{align}
where $\hat t=-t(\delta_{n_1, n_1'+1} +\delta_{n_1, n_1'-1})$ and
$\varepsilon(k_y,k_z)=-2t(\mathrm{cos}(k_y a)+\mathrm{cos}(k_z a))$. To apply
DMFT to the transition regime, we must calculate the local component of the
Green's function at each site and map each onto an auxiliary impurity.

We can rewrite Eq.~\eqref{G_mit} in a block matrix divided into the three
regimes $\cM$, $\cT$ and $\cI$, that is,
\begin{equation}\label{G_mat}
  [G(k_y,k_z;\iw)]_{n_1 n_1'}=\left[
    \begin{array}{c c| c c c |c c}
      F_{\cM} & & & & & &\\
      & & t & & & & \\ \hline
      & t & & & & \\ 
      & & & F_{\cT} & & & \\ 
      & & & & & t & \\ \hline
      & & & & t & & \\
      & & & & & & F_{\cI} \\
    \end{array}\right]^{-1}
\end{equation}
where we define the three block matrices by
\begin{align*}
[F_{\cM}]_{n_1 n_1'}=&\underbrace{(\iw + \mu - \varepsilon(k_y,k_z)- \Sigma_{\cM}(\iw))}_{\equiv z_{\cM}}\delta_{n_1 n_1'} - t_{n_1 n_1'}\\
[F_{\cT}]_{n_1 n_1'}=&\underbrace{(\iw + \mu - \varepsilon(k_y,k_z)- \Sigma_{n_1}(\iw))}_{\equiv  z_{n_1}}\delta_{n_1 n_1'} - t_{n_1 n_1'}\label{z_T} \\
[F_{\cI}]_{n_1 n_1'}=&\underbrace{(\iw + \mu - \varepsilon(k_y,k_z)- \Sigma_{\mathrm{ins}}(\iw))}_{\equiv  z_{\cI}}\delta_{n_1 n_1'} - t_{n_1 n_1'}.
\end{align*} 
Note that $z_{\cM}$ and $z_{\cI}$ are fixed while $z_{n_1}$ varies across the
sites.

Using block matrix inversion
\begin{equation}\label{inv_rel}
  \left.\left[
    \begin{array}{c| c  }
      \mathbf{A} &  \mathbf{B}  \\ \hline
      \mathbf{C} &  \mathbf{D}   
    \end{array}\right]^{-1}\right|_{\in \mathbf{A}}
  =[\mathbf{A} - \mathbf{B} \mathbf{D}^{-1}\mathbf{C}]^{-1}
\end{equation}
we obtain the complete form of Green's function in the transition regime $\cT$
(a $N\times N$ matrix) into which all the degrees of freedom of metallic and
insulating regions are incorporated:
\begin{widetext}
\begin{align}\label{G_inv}
  [G(k_y,k_z)]|_{n_1,n_1'\in \cT}&= \big[ \underbrace{[F_\cT]}_{\mathbf{A}}
    -(\underbrace{\hat t_{\cT \cM}[F_\cM]^{-1}\hat t_{\cM \cT}+\hat t_{\cT \cI}[F_\cI]^{-1}\hat t_{\cI \cT}}_{\mathbf{B} \mathbf{D}^{-1}\mathbf{C}} \big]^{-1} \nonumber \\
  &=\left[\begin{matrix}
      z_{11}-t^2 R_{\cM} &  t  &    &   & & 0 \\
      t  &  z_{22}  & t  &   \\
      & t   &  \ddots  & \ddots  \\
      &     &  \ddots & \ddots & t  \\
      &   &  &  t & z_{N-1,N-1} &t  \\
      0      &  & &   &   t & z_{NN} - t^2R_{\cI}
    \end{matrix}\right] ^{-1}
\end{align}
\end{widetext}
where $R_{\cM}\equiv [F_{\cM}^{-1}]_{00}$, $R_{\cI}\equiv
[F_{\cI}^{-1}]_{N+1,N+1}$ and $\hat t_{\cT\cM(\cI)}$ is the overlap between
$\cT$ and $\cM(\cI)$. The effect of integrating out the degrees of freedom in
$\cM$ and $\cI$ is captured by $t^2 R_{\cM}$ and $t^2 R_{\cI}$ at the $(1,1)$
and $(N,N)$ components respectively.

To compute $R_{\cM}$ and $R_{\cI}$, we again rely on Eq. \eqref{inv_rel}. Since
$[F_{\cM}]$ takes a symmetric tridiagonal matrix form equal to
\begin{align}
  F_{\cM} =\left[\begin{matrix}
      z_{\cM}  &  t  &    &   \\
      t  &  z_{\cM}  & t  &   \\
      & t   &  z_{\cM} & \ddots  \\
      &     & \ddots    & \ddots  
    \end{matrix}\right] 
  = \left[\begin{array} {c| c c c}
      z_{\cM}  &  t  &           &   \\ \hline
      t        &     &           &   \\
      &        &  F_{\cM} &   \\
      &        &          & 
    \end{array}\right]
\end{align}
we see the matrix repeats itself inside. As a direct consequence of
\eqref{inv_rel}, we obtain the following recursive equation:
\begin{equation}\label{eq_F}
  [F_{\cM}^{-1}]_{00}= R_{\cM}=\frac{1}{z_{\cM}-t^2R_{\cM}}
\end{equation}
where the solution is
\begin{equation}
  R_{\cM}=  \frac{z_{\cM} -\sqrt{(z_{\cM})^2-1}}{t}.
\end{equation}
$R_{\cI}$ is obtained by the same procedure.

Finally, we need to convert the mixed representation form \eqref{G_mat} into
the pure real-space representation. Performing the inverse Fourier
transformation with respect to $k_y$ and $k_z$, we can obtain the local Green's
function at the site $n_1$
\begin{align} \label{inv_FT}
  [G]_{n_1 n_1}&=\int \frac{d^2k}{(2 \pi)^2} [G(k_y,k_z)]_{n_1n_1}\nonumber \\
  &=\int d\varepsilon [G(\varepsilon)]_{n_1 n_1} D^{2D}(\varepsilon)
\end{align}
where the $\varepsilon$ dependence of $G$ comes from
$\varepsilon=\varepsilon(k_y,k_z)$. Here, $D^{2D}(\varepsilon)$ is the density
of states of non-interacting 2D square lattice whose analytic expression is
known and the integration \eqref{inv_FT} is performed numerically.

\end{document}